\documentclass[twocolumn,showpacs,amsmath,amssymb,prl]{revtex4}

\usepackage{graphicx}
\usepackage{dcolumn}
\usepackage{bm}

\begin{document}

\title{Efficient Monte Carlo simulation of a glass forming binary mixture}

\author{E. Flenner}
\author{G. Szamel}
\affiliation{Department of Chemistry, Colorado State University, Fort. Collins,
CO 80523}

\begin{abstract}
We propose and use a novel, hybrid Monte Carlo algorithm that combines 
configurational bias particle swaps with parallel tempering. We use this new method to
simulate a standard model of a glass forming binary mixture above and below the
so-called mode-coupling temperature, $T_{MCT}$. 
We find that an \textit{ansatz} that was used previously to extrapolate 
thermodynamic quantities to temperatures below $T_{MCT}$ breaks down in the vicinity 
of the mode-coupling temperature.
Thus, previous estimates of the so-called Kauzmann temperature need to be reexamined. 
Also, we find that the Adam-Gibbs relations $D \propto \exp(-a/TS_c)$ and 
$\tau \propto \exp(b/TS_c)$, which
connect the diffusion coefficient $D$ and the relaxation time $\tau$ 
with the configurational entropy 
$S_c$, are valid for all temperatures for which the configurational and
vibrational contributions to the free energy decouple.
\end{abstract}

\date{\today}


\maketitle

Understanding the nature of the glass transition has been of great interest 
for several decades. One of the earliest paradigms \cite{Kauzmann,AdamGibbs}, 
that has recently been reformulated \cite{MezardParisi} and subsequently received
considerable attention, assumes the existence of an ``ideal'' glass transition 
which occurs when the entropy of the supercooled liquid becomes equal to the
entropy of a disordered solid. 
This paradigm has stimulated several simulational studies \cite{Sciortino,Coluzzi}
which have confirmed both the 
qualitative description of supercooled liquids' dynamics and even the quantitative, 
numerical prediction of the transition temperature,
the so-called Kauzmann temperature $T_K$, 
for a popular model of a glass forming binary mixture. However, 
these simulational investigations were restricted to
temperatures above the so-called mode-coupling temperature, $T_{MCT}$. Thus, in these
previous studies, 
in order to investigate the equality of the liquid and disordered solid 
entropies, one had to extrapolate higher temperature data obtained directly from
simulations to significantly lower temperatures (for example, for the binary mixture 
studied in Refs.~\cite{Sciortino,Coluzzi}
it was found that $T_{MCT}/T_K \approx 1.45$). The commonly used 
extrapolations were questioned in a recent investigation 
of the configurational entropy \cite{Yan}. This study is unique in that
it accessed directly the low temperature region below $T_{MCT}$. It 
showed that the commonly used 
extrapolations break down below $T_{MCT}$
and the existence of the Kauzmann temperature was put in doubt. Reference \cite{Yan} 
used a novel density-of-states Monte Carlo method. Also, it used 
a binary mixture model that was somewhat different \cite{AndersenPNAS}
than the model used in most of the previous simulations. Finally,
although several system sizes were considered in Ref.~\cite{Yan}, the largest system
(216 particles) was significantly smaller than systems used in previous studies 
(the largest ones had 1000 particles). 

The goal of our investigation was to address the question of the existence of the 
Kauzmann temperature for the original model studied in 
Refs.~\cite{Sciortino,Coluzzi}. 
Since the long relaxation times
below $T_{MCT}$ makes equilibration of molecular 
and Brownian dynamics simulations very difficult, we propose and use a novel,  
specialized Monte Carlo algorithm designed to
decrease the time between independent measurements of strongly supercooled liquids.
This new method allows us to 
obtain accurate thermodynamic quantities for temperatures below the mode-coupling
temperature.  
We find that the previously used extrapolations of
thermodynamic quantities to estimate the Kauzmann
temperature break down in the vicinity of the mode-coupling temperature.
Furthermore, by using the results of Brownian dynamics 
simulations, we demonstrate that the Adam-Gibbs relations \cite{AdamGibbs}
are valid for low temperatures.  We propose using the Adam-Gibbs
relation and the results of the Monte Carlo simulation to predict the
diffusion coefficient.

To directly access temperatures below the mode-coupling temperature,
we combined non-local trial moves with parallel
tempering. Non-local trial moves, identity exchanges, have been 
effective in speeding up simulations with 
different size particles \cite{Grigera,Faller}.
However, high density and large size disparity drastically reduce the acceptance 
rate of identity exchanges \cite{Faller}, which decreases their 
usefulness in dense glass forming liquids at low temperatures. 
To overcome this, we used parallel tempering,
which attempts to exchange replicas of the system which
are being simulated at different temperatures.  This allows 
configurations which are trapped in a deep potential energy
minimum at low temperatures to change considerably
by being simulated at higher temperatures.  A 
variation of this technique, which combines replica exchange 
with molecular dynamics, 
has been shown to speed up the equilibration of the
system studied in this work \cite{KobRE}. 

We simulated an 80:20 binary mixture 
introduced by W. Kob and H. Andersen \cite{KobAndersen}.  The
interaction potential is 
$V_{\alpha \beta}(r_{ij}) = 4 \epsilon_{\alpha \beta} 
\left[ (\sigma_{\alpha \beta}/r_{ij})^{12} -
(\sigma_{\alpha \beta}/r_{ij})^{6} \right]$ where 
$\alpha, \beta \in \{A,B\}$, $\epsilon_{AA} = 1.0$, 
$\epsilon_{AB} = 1.5$, $\epsilon_{BB} = 0.5$, $\sigma_{AA} = 1.0$, 
$\sigma_{AB} = 0.8$, and $\sigma_{BB} = 0.88$.  The results
are presented in reduced units with $\epsilon_{AA}$ and $\sigma_{AA}$ 
being the units of energy and length, respectively.  We simulated 1000
particles in a fixed cubic box with a box length of 9.4.
We performed a parallel tempering 
Monte Carlo simulation and a series of Brownian dynamics simulations.  
The Monte Carlo simulation was 
performed using the following set of 
temperatures, $T=0.62$, 0.59, 0.56, 0.53, 0.50, 0.48, 0.46, 0.44,
0.43, 0.42, 0.41, and 0.40.  The Brownian dynamics simulations were performed
at $T = 5.0$, 3.0, 2.0, 1.5, 1.0, 0.9, 0.8, 0.6, 0.55, 0.5, 0.47, 0.45, and 0.44.  

The details and results of the 
Brownian dynamics simulation have been presented elsewhere \cite{SF,FS1}. Here we briefly
describe the Monte Carlo simulation. 
We utilize three trial moves: a standard, local 
single particle displacement, a configurational bias particle
swap, and parallel tempering.  
The configuration bias particle exchange attempts to 
swap two particles of different sizes.  
The smaller particle will always fit in the space left by 
the larger particle, but, since the density is high, 
the converse is rarely true.  To increase
the acceptance rate, 50 trial configurations are explored for the larger particle
around the former position of the smaller particle.  One of the
trial positions is chosen with a probability which depends on the 
potential energy, and the move is accepted with a probability so
that detailed balance is maintained \cite{Faller,FrenkelSmit}.  The 
configurational bias increases the acceptance rate, but the acceptance rate 
for the particle swaps is still very small, around $10^{-8}$ at $T=0.5$,
which is the lowest temperature in which the swaps were attempted.  
It has been shown that identity exchange
can decrease the equilibration time of a simulation dramatically \cite{Grigera,Faller}
even if the acceptance rate is very small.
Parallel tempering consists
of an attempted exchange of particle positions 
between adjacent temperatures
\cite{FrenkelSmit}.  Thus, the configurational bias particle swaps
can decrease the correlation time for the configurations which begin
the simulation at the temperatures in which the swaps
are not attempted.  

To estimate the efficiency of our algorithm, we compared the energy
correlation time measured in Monte Carlo moves per particle with the
energy correlation time in the Brownian Dynamics simulation measured in 
Brownian Dynamics time steps. We found that the former time increases slower
with decreasing temperature than the latter time. 
The energy correlation time for the Monte Carlo 
simulation at $T=0.4$ corresponds to generating one 
statistically independent configuration approximately every five days 
for a parallel algorithm using four threads
on a hyperthreaded dual-processor 3.2 GHz Pentium workstation. 

\begin{figure}
\includegraphics[scale=0.3]{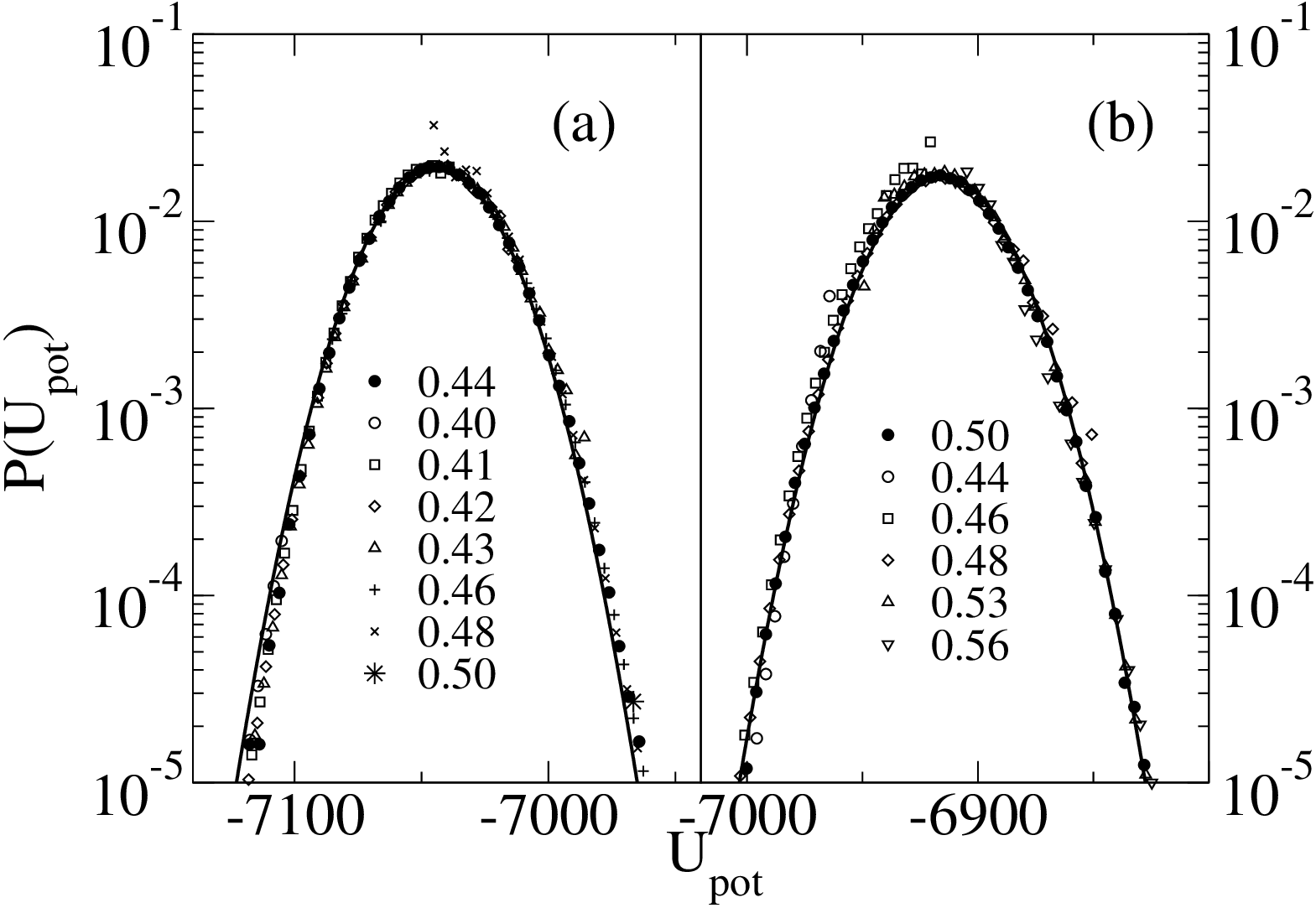}
\caption{\label{reweight1}The re-weighted potential energy probability distributions.  
The solid
line is a Gaussian distribution  
$\sqrt{1/(2 \pi k_B C_{V,pot}(T) T^2)}
\exp(-(U_{pot} - \left< U_{pot} \right>)^2/(2 k_B C_{V,pot}(T) T^2)$
for $T = 0.44$ and $T = 0.50$.}
\end{figure}
\begin{figure}
\includegraphics[scale=0.3]{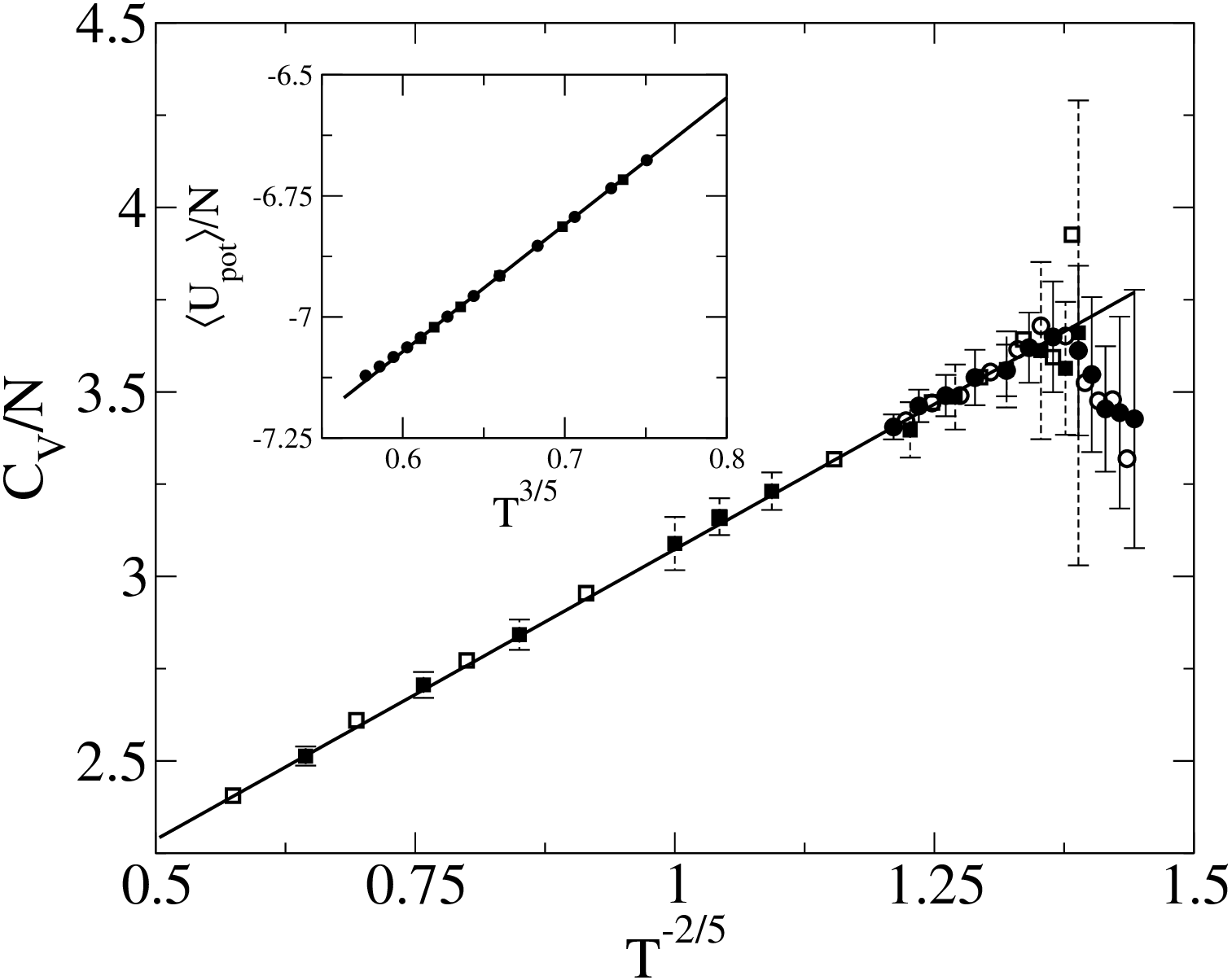}
\caption{\label{spheat}The specific heat per particle as a function of $T^{-2/5}$ 
calculated from energy fluctuations
(closed symbols) and the derivative of the energy (open symbols) for the
Brownian dynamics (squares) and the Monte Carlo (circles).  The solid line
is $C_V/N = 0.6 a T^{-2/5} + 1.5$ where $a$ was obtained from fitting the average 
potential energy $\left<U_{pot}\right>$
to a function of the form $a T^{3/5} + b$ (the $a$ and $b$ parameters
obtained from the fit are $a=-8.6547$ and $b=2.6362$). Inset: the average potential 
energy per particle as a function of $T^{3/5}$; the solid line is a $a T^{3/5} + b$ fit.}
\end{figure}

We used a variety of different checks for equilibration: monitoring 
the running average of the potential energy
$U(N) = (1/N) \sum_{i=1}^N U_i$ where $U_i$ is the potential energy for step $i$
($U(N)$ does not show any systematic drift), comparing specific heat calculated using
the derivative of the energy and energy fluctuations 
(they agree, see Fig.~\ref{spheat}), comparing the temperature assumed in the
simulation algorithm with the so-called 
configurational temperature \cite{Jepps} (they agree),
\textit{etc.} \cite{equilibrate}. Here we discuss in some detail a stringent
equilibration test introduced by Yamamoto and Kob \cite{KobRE}.
This so-called re-weighting 
procedure relies upon the fact that an equilibrium distribution of the energy 
for a temperature $T_1$ can be determined
from an equilibrated simulation at a different temperature $T_0$.
Shown in Fig.~\ref{reweight1}a and Fig.~\ref{reweight1}b  
are the re-weighted probability distributions of the potential energy for
$T = 0.44$, and 0.5, respectively.  The filled circles in the figures
are the probability distributions calculated from simulation runs
at $T=0.44$ and 0.5, respectively.
The other symbols are the re-weighted probability distributions, and
the solid lines are Gaussian distributions determined from the average 
energies and the specific heats calculated at the respective temperatures. 
Note that there is a slight systematic deviation from a Gaussian
distribution at the lowest temperatures. This deviation is within the
uncertainty of the calculation, thus it is not clear whether it has any
significance. In any case, it should be noted that the re-weighted probability 
distributions still superimpose very well. Thus, at all temperatures  
there is good overlap between the re-weighted distributions and this fact provides 
strong evidence that the Monte Carlo simulation has properly equilibrated.  

The specific heat calculated from energy fluctuations and from 
the derivative of the energy determined from the Brownian dynamics and
the Monte Carlo simulation is shown in Fig.~\ref{spheat}.  Note that
a correction due to the finite simulation time \cite{comment,Muller,Ferrenberg} 
has been applied to the
specific heat calculated from the energy fluctuations.  The agreement 
between the Monte Carlo, the Brownian dynamics, and the two methods 
of calculating the specific heat is very good except for the
derivative of the energy for the Brownian dynamics simulation at the lowest 
temperature. There is a peak in the specific heat
around $T \approx 0.45$, which is close to the usually cited mode-coupling temperature
$T_{MCT} \approx 0.435$ \cite{KobAndersen} determined  
from fits to the diffusion coefficient and the relaxation time. 
It is clear from Fig.~\ref{spheat} that the extrapolation that was used in prior
studies \cite{Sciortino,Coluzzi} is violated below the mode-coupling temperature.
A similar violation was observed in a recent investigation of a different binary mixture 
\cite{Yan}. Recall that the former studies only simulated systems at 
temperatures higher than $T_{MCT}$, whereas the latter one was able to access 
temperatures below $T_{MCT}$.

\begin{figure}
\includegraphics[scale=0.42]{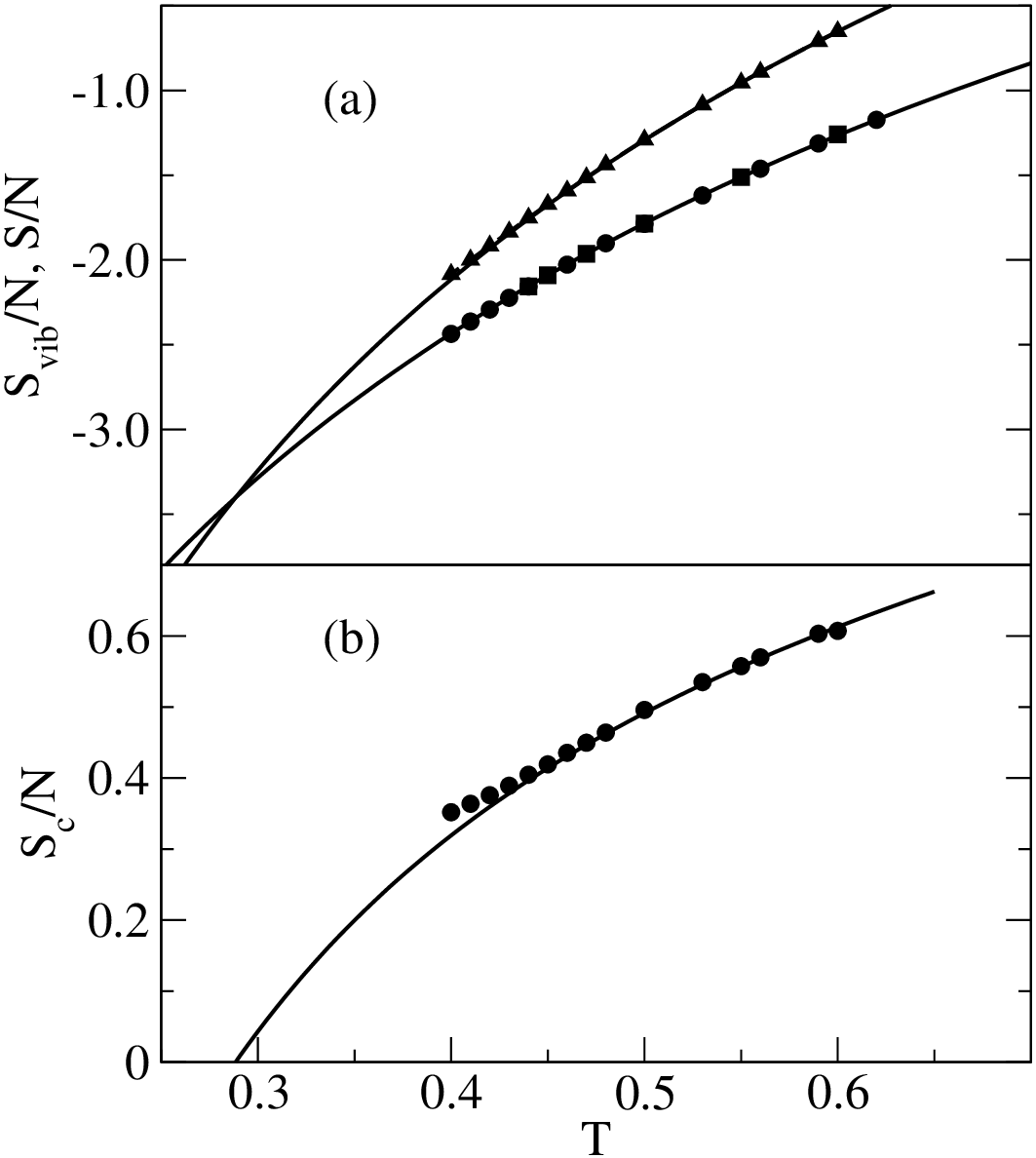}
\caption{\label{entropy}(a) The total entropy per particle $S/N$ (triangles) and 
the disordered solid entropy per particle $S_{vib}/N$. The latter entropy has been 
calculated using the inherent structures obtained from the Monte Carlo (circles) 
and the Brownian dynamics (squares)
simulations.  
The solid lines are the commonly used extrapolations described in the text. 
(b) The configurational entropy per particle, $S_c/N = S/N - S_{vib}/N$. The solid line
is the extrapolation obtained from the difference of the extrapolations shown in (a).
The error bars for $S/N$, $S_{vib}/N$, and $S_c/N$ are smaller than the size 
of the symbols.}
\end{figure}

In Fig.~\ref{entropy}a we show the total liquid entropy, 
$S$ and the so-called disordered solid
entropy, $S_{vib}$.
To find the total entropy $S$ we performed a standard thermodynamic
integration along the $T = 5.0$ isotherm and then along the $V_0 = (9.4)^3$ isochore.
For $T \ge 0.62$ we utilized a commonly used fit for the specific heat
shown as a solid line in Fig.~\ref{spheat}. 
For $T < 0.62$ we numerically integrated the specific heat.
The results of the numerical integration are shown as triangles in Fig.~\ref{entropy}a.
The solid line going through the triangles for $T> 0.45$ and slightly 
deviating from them for $T< 0.45$ is the standard extrapolation \cite{Sciortino,Coluzzi}
that relies on using the fit shown as a solid line in Fig.~\ref{spheat} 
for \emph{all} temperatures. 

To evaluate the disordered solid entropy we followed the procedure used in 
previous studies\cite{Stillinger,Nave,Sciortino,Scala,Speedy,Sastry}.
First, we determined the inherent structures by quenching 500-1000
configurations. 
Then we checked that for $T \le 0.62 $ the system can be described by a configurational 
and a vibrational part \cite{Sciortino}. 
We calculated the
entropy of the disordered harmonic solid  
by diagonalizing the Hessian matrix calculated
at the inherent structures and determined the vibrational
frequencies $\omega_i$. 
Next, we calculated the vibrational contribution to the entropy
$S_{vib} = \left< \sum_{i=1}^{3N - 3} [1 - \ln(\beta \hbar \omega_i)] \right>^{\prime}$
where $< \cdot >^{\prime}$ denotes an average over the
inherent structures (note that throughout this paper 
we set Planck's constant equal to one). The results are shown as 
circles and squares in Fig.~\ref{entropy}a.
The solid line going through the circles and squares is obtained by
fitting 
$ \left< \sum_{i=1}^{3N - 3}\ln(\omega_i)] \right>^{\prime}$
to a polynomial in $T$ of degree 2. This quantity, which is the contribution 
to $S_{vib}$ that originates from the the vibrational
frequencies, is almost temperature-independent.

In Fig.~\ref{entropy}b we show the configurational entropy
$S_c = S - S_{vib}$. Note that $S_c$ was only calculated for 
temperatures for which the the system can be divided into 
a configurational and a vibrational part, \textit{i.e.} for $T \le 0.62$.
In Fig.~\ref{entropy}b we also compare our results with the previously used extrapolation 
of the configurational entropy. 
This extrapolation results in the Kauzmann temperature $T_K = 0.29$, which is 
very close to previous estimates \cite{Sciortino,Coluzzi}.
Note that below the mode-coupling temperature our results deviate from this
extrapolation. 
Thus, the previous estimates of the Kauzmann temperature
have to be reexamined. 
The peak in the specific heat 
indicates that the total entropy is larger than the previous
estimate, and that the Kauzmann temperature,
if it exists at all, is lower. Our results are consistent with 
those obtained in Ref.~\cite{Yan} for a 
different binary mixture.  
 
\begin{figure}
\includegraphics[scale=0.3]{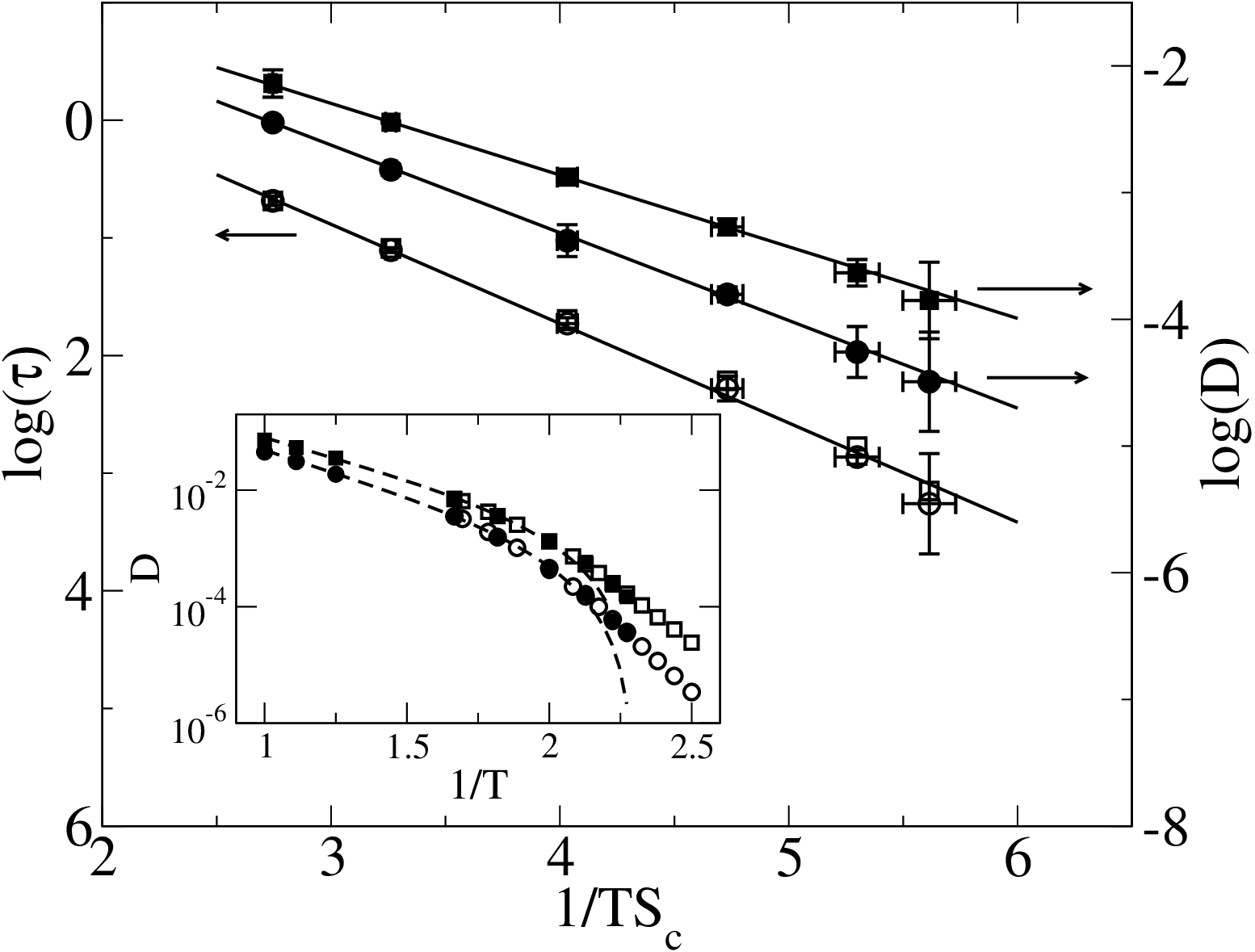}
\caption{\label{adamgibbs}Test of the Adam Gibbs relations
$D \propto \exp(-a/TS_c)$ (solid symbols and right vertical axis) 
and $\tau \propto \exp(b/TS_c)$ (open symbols and left vertical axis) 
for the A (circles) and the B particles (squares).
Inset: The diffusion coefficient determined directly from
the Brownian dynamics simulations (closed symbols) and predicted using the 
Adam-Gibbs relation (open symbols) for the A (circles) 
and B particles (squares). The Adam-Gibbs prediction uses $S_c$ determined 
from the Monte Carlo simulations. The dashed curves are mode-coupling theory
power law fits $a(T-0.435)^\gamma$ over the temperature range 
$0.8 \le T \le 0.5$.}
\end{figure}
We examined the Adam-Gibbs relations $D \propto \exp(-a/TS_c)$ and 
$\tau \propto \exp(b/TS_c)$ \cite{AdamGibbs}.
Shown in Fig.~\ref{adamgibbs} are the logarithms of the diffusion coefficients
and the relaxation times (the relaxation times are obtained from the 
self-intermediate scattering function in the usual way \cite{KobAndersen,SF})
plotted as a function of $1/TS_c$ for $T \le 0.6$.  
The diffusion coefficients and the relaxation times are 
determined from the Brownian dynamics simulations.
The configurational entropy is determined from the Monte
Carlo simulation and the Brownian dynamics simulations.
Although some curvature can be seen in the relaxation time data,
the fits are very good and verify the Adams-Gibbs relation for this
temperature range.  Assuming that the Adam-Gibbs relation holds 
at lower temperatures, we can predict the diffusion coefficient down to 
$T = 0.4$ from the results of the Monte Carlo simulation; see the inset in 
Fig.~\ref{adamgibbs}.

In summary, we propose and use a novel, hybrid Monte Carlo
algorithm which allows accurate calculation of equilibrium quantities
below the mode-coupling temperature.  Our method combines non-local configuration
bias particle swaps and parallel tempering.  The results obtained
with this algorithm put in doubt the commonly used extrapolation of the supercooled
liquid entropy and previous estimates of the Kauzmann temperature. Moreover, we
show that the Adam-Gibbs relations $D \propto \exp(-a/TS_c)$ and 
$\tau \propto \exp(b/TS_c)$ hold for all temperatures for which the configurational and
vibrational contributions to the free energy decouple.

We gratefully acknowledge the support of NSF Grants No.~CHE 0111152 and
CHE 0517709.


\begin{thebibliography}{99}
\bibitem{equilibrate}  
We also checked that the energy correlation time did not change appreciable 
for over half the simulation time and the specific heat could be calculated 
consistently from smaller segments of the simulation.
\end{thebibliography}
\end{document}